\input{psfig}
\documentclass{aa}
\usepackage{psfig}
\thesaurus{07.13.2; 07.03.1; 07.16.1; 07.19.2}
\title{Kuiper Belt evolution due to dynamical friction}
\author{A. Del Popolo\inst{1,2} \and ~E. Spedicato\inst{2} \and M. Gambera\inst{1,3}}  
\offprints{A. Del Popolo (antonino@astrct.ct.astro.it); E. Spedicato ( emilio@ibguniv.unibg.it); M. Gambera (mga@sunct.ct.astro.it)}
\institute{Istituto di Astronomia dell'Universit\`a di Catania, 
Viale A.Doria, 6 - I 95125 Catania, ITALY \and
Dipartimento di Matematica, Universit\`{a} Statale di Bergamo, Piazza Rosate, 2 - I 24129 Bergamo, ITALY
\and Osservatorio Astrofisico di Catania and CNR-GNA, 
Viale A.Doria, 6 - I 95125 Catania, ITALY 
}
\titlerunning{Kuiper Belt evolution due to dynamical friction}
\authorrunning{Del Popolo et al.}
\date{}
\begin{document}
\maketitle
\begin{abstract}
In this paper we study the role of dynamical friction on the
evolution of a population of large objects ($m>10^{22}$ g) at heliocentric
distances $>70$ AU in the Kuiper Belt. We show that the already flat
distribution of these objects must flatten further due to 
non-spherically symmetric distribution of matter in the Kuiper Belt.
Moreover the dynamical drag, produced by dynamical
friction, causes objects of masses
$\geq 10^{24} {\rm g}$ to lose angular momentum
and to fall
through more central regions in a timescale $\approx ~10^9 {\rm yr}$.
This mechanism
is able to transport inwards objects of the size of Pluto, supposing
it was created beyond $~50 {\rm AU}$, according to a Stern \& Colwell's (1997b)
suggestion.
\end{abstract}

\section{Introduction}

The current model for comets in the solar system supposes that a vast cloud of
cometary objects orbits the Sun. This cloud consists of three components.
The inner one, referred to as the Kuiper Belt (hereafter KB) (Edgeworth 1949; Kuiper 1951),
is a disc like structure of $\geq 10^{10}$ comets extending from $40 - 10^3 {\rm AU}$
from the Sun (Weissman 1995; Luu et al. 1997). The KB has been
proposed as the source of the Jupiter-family short-period (hereafter SP) 
comets. The
second component, referred to as the Oort inner cloud, or the Hills cloud (Hills
1981), is supposed to be a disc, thicker than KB, containing $%
10^{12}- 10^{13}$ objects lying $\sim 10^3 - 2\times 10^4 {\rm AU}$
from the Sun. It
is proposed as a source of long-period (hereafter LP) and Halley-type SP comets
(Levison 1996). The last component, the Oort cloud (Oort 1950), is a
spherical cloud of $10^{11} - 10^{12}$ cometary objects with nearly isotropic
velocity distribution extending from $2\times 10^4$ to $2\times 10^5 {\rm AU}$.
Even if the Oort cloud was considered in the past as the fundamental reservoir
of LP comets which have been  brought into the inner solar system by perturbations
due to the galactic tidal field, molecular clouds and passing stars,
nowadays it has been shown that it can contribute only for a small part
to the LP population of comets (Duncan et al. 1988; Wiegert \& Tremaine 1997).\\
Observational confirmation of the KB was first achieved with
the discovery of object 1992QB1 by Jewitt \& Luu (1993). To date
over 40 KB objects (hereafter KBO) with diameters between 100 and 400 km
have been discovered
and the detection statistics obtained suggest
that a complete ecliptic survey would reveal $~ 7 \times 10^4$ such
bodies orbiting between 30 and $50 {\rm AU}$.
Such a belt of distant icy planetesimals
could be a more efficient source of SP comets than the Oort cloud (Fernandez
1980; Duncan et al. 1988). Dynamical simulations have shown that a cometary
source with low initial inclination distribution was more consistent with
the observed orbits of SP comets than the randomly distributed inclinations
typical of the comets in the Oort cloud (
Quinn et al. 1990; Levison \& Duncan 1993).
According to other simulations the greatest part of objects in the KB
should be stable for the age of the solar system. However if the population of
the KB is $\sim 10^{10}$ objects, weak gravitational perturbations
provide a large enough influx to explain the current population of SP comets
(Levison $\&$ Duncan 1993; Holman $\&$ Wisdom 1993; Duncan et al. 1995). In
particular Levison $\&$ Duncan (1993) and Holman $\&$ Wisdom (1993), studied
the long term stability of test particles in low-eccentricity and
low-inclination orbits beyond Neptune, subject only to the gravitational
perturbations of the giant planets. They found orbital instability on
timescales $<10^7 {\rm yr}$ interior to $ 33  -  34 {\rm AU} $,
regions of stability and
instability in the range $ 34  - 43 {\rm AU}$
and stable orbits beyond $43 {\rm AU}$.\\
A study by
Malhotra (1995a) showed that the KB is characterized by a highly non
uniform distribution: most of the small bodies in the region between Neptune
and $50 {\rm AU}$ would have been swept into narrow regions of orbital resonance
with Neptune (the 3:2 and 2:1 orbital resonances, respectively located at
distances from the Sun of $39.4 {\rm AU}$ and $47.8 {\rm AU}$). The orbital inclinations 
{\it i} of
many of these objects would remain low ($i<10^{\rm o}$) but the eccentricities
{\it e} would have values from 0.1 to 0.3. At the same time many of the
trans-neptunian objects discovered lie in low-inclination orbits, as
predicted by the dynamical models of Holman \& Wisdom (1993) and
Levison \& Duncan (1993). A more detailed analysis of this distribution
reveals that most objects inside $~ 42 {\rm AU}$ reside in
higher-$e$, $i$ orbits locked in mean motion resonance with Neptune, but
most objects beyond this distance reside in non-resonant orbits
with significantly lower eccentricities and inclinations.  \\
After the previously quoted discovery of $100  -  200$ km sized objects (Jewitt $\&$ Luu
1993; Jewitt \& Luu 1995; Weissmann \& Levison 1997),
proving that the KB is populated,
Cochran et al. (1995) have reported Hubble Space Telescope
results giving the first direct evidence for comets
in the KB. Cochran's observations imply
that there is a large population ($>10^8$) of
Halley-sized objects (radii $\sim 10$ km) within $\sim 40 {\rm AU}$ of the Sun, made
up of low inclination objects ($i<10^{\rm o}$). At the time of the
publication, Cochran's et al. (1995) results were criticized on two grounds:\\
1) the detections were statistical in nature, and the authors were not able to
fit orbits to their objects;\\
2) the number of detections did not agree with extrapolation of the size
distribution of large KBOs determined from early ground-based observations
(but Weissman \& Levison ~1997 showed that Cochran's et al. 1995 results
were in agreement with the number of KBOs needed to populate the
Jupiter-family comets);
moreover Brown et al. 1997 contended that detections reported in
Cochran et al. (1995) were not possible, based on an analysis of the noise
properties of the data. In a recent paper, Cochran et al. 1998, confirmed
the early results by means of a new analysis.\\
The spatial dimensions and mass distribution in the KB are
poorly known. Yamamoto et al. (1994) have applied a planetesimal model
to the trans-neptunian region, finding that the maximum number density of
the planetesimal population should be about at $100  -  200 {\rm AU} $ and the
planetesimal disc itself can extend up to distances $\approx 10^3 {\rm AU}$.
This is in agreement with detection by IRAS of discs around main sequence
stars, Vega (Aumann et al. 1984), $\beta$ Pictoris (Smith \& Terrile 1984),
extending to several hundred AU. From the available radio and infrared
data, Beckwith \& Sargent (1993) conclude that disc masses may range
between $10^{-3} M_{\odot}$ to $1 M_{\odot}$ and extend from a
few hundred AU to more than $10^3 {\rm AU}$. In short both theoretical arguments
and observations strengthen the view that our solar system is surrounded by
a flattened structure of planetesimals, extending perhaps to several
hundred AU. \\
Although originally it had been thought that the population might
be collisionless, recent work (Stern 1995) has shown that the collisional
effects cannot be neglected over $4.5$ Gyr. As shown
by Stern (1995, 1996a,b) and 
Stern \& Colwell (1997a,b), the
collisional evolution is an important evolutionary process in the disc
as a whole, and moreover, it is likely to be the dominant evolutionary
process beyond $~ 42 {\rm AU}$. In the case of larger planetesimals the evolution
is connected to the energy loss due to dynamical friction, which transfers
kinetic energy from the larger planetesimals to the smaller ones. This
mechanism, in the early solar system,
provides an energy source for the small planetesimals that is comparable
to that provided by the viscous stirring process (Stewart \& Wetherill 1988;
Weidenschilling et al. 1997).\\
The objective of this paper is to examine the role of dynamical friction,
in the primordial KB, 
in the orbital evolution of the largest planetesimals that lie at a
heliocentric distance $>70 {\rm AU}$ (at this distance the effects of the planets
decline
rapidly to zero
and
only a small fraction of objects is influenced by planetary 
perturbations -  Wiegert \& Tremaine 1999; Stern \& Colwell 1997b).
While the importance of dynamical
friction in planetesimal dynamics was demonstrated in several papers, 
(Stewart \& Kaula 1980; Horedt 1985; Stewart \& Wetherill 1988)
and in particular in the case of the planetary accumulation
process, the role of this effect on the orbital
evolution of the largest planetesimals and the consequent
change of mass distribution in KB was never studied.\\
The paper is organized as follows. In Sect. 2, we review the role of
encounters and collisions in KB.
In Sect. 3, we introduce the equations
to calculate dynamical friction effects. In Sect. 4 we
describe how we use these
equations to determine the evolution of the largest bodies population in
KB. In Sect. 5 we discuss the results of the calculation and 
we also show (supposing the scenario proposed
by Stern \& Colwell 1997b of Pluto formation beyond $~50 {\rm AU}$ to be correct)
how dynamical friction is able to transport an object
of the size of Pluto from $~50 {\rm AU}$ to the actual position. In Sect. 6 we give
our conclusions.

\section{Encounters and collisions in KB}

As shown in several N-body simulations (Stern 1995), the structure of KB
in the region $30  -  50 {\rm AU}$ is fundamentally due to two processes: 
\begin{itemize}
\item 1) dynamical erosion due to resonant interactions with Neptune
(Holman \& Wisdom 1993; Levison \& Duncan 1993; Duncan et al. 1995;
Malhotra 1995a).
In this region
the KB has a complex
structure. Objects with perihelion distances $\simeq 35 {\rm AU}$ are unstable. For
orbits with $e \geq 0.1$ and with semi-major axis $a< 42 {\rm AU}$ the only stable orbits are those in
Neptune resonances. Between $40 {\rm AU}$ and $42 {\rm AU}$ at low inclinations and 
between $36 {\rm AU}$ and $39 {\rm AU}$ with $i \simeq 15^o$ the orbits are unstable.
\item 2) collisional erosion (Stern 1995, 1996a,b; Stern \& Colwell 1997a,b).
Starting from a primordial disc having a mass of $40 M_{\odot}$ collisions
are able to reduce its mass to $0.1  -  0.3 M_{\odot}$ in $10^9 {\rm yr}$ if
$\langle e \rangle \geq 0.1$.
\end{itemize}
Then the $30 - 50 {\rm AU}$ zone is both collisionally
and dynamically evolved, since dynamics acted to destabilize most orbits
with $a<42 {\rm AU}$ and were able to induce eccentricities that caused collisions out to
almost $50 {\rm AU}$. The dynamically and collisionally evolved zone might extend
as far as $\simeq 63 {\rm AU}$, if Malhotra's (1995a) mean motion resonance
sweeping mechanism is important. Beyond
this region one expects there to be a collisionally evolved zone where accretion
has occurred but eccentricity perturbations by the giant planets have been too low
to initiate erosion. Beyond that region it is expected a primordial zone
in which the accretion rates have hardly modified the initial population
of objects. Supposing that the last assumption is correct 
and that the radial distribution of heliocentric surface
mass density in the disc, $\Sigma(R)$, can be described by
$\Sigma(R) \propto R^{-2}$ (Tremaine 1990; Stern 1996a,b) and supposing
that in the zone $ 30  -  50 {\rm AU}$ of the primordial disc
$40  -  50 M_{\oplus}$ of matter was present, we expect $\simeq 35 M_{\oplus}$ in the zone
$70 - 100 {\rm AU}$ of the present KB. 
As the effects of planets are negligible for planetesimals
beyond $70 {\rm AU}$,
the orbital motion
there can be considered not far from Keplerian and circular
(Brunini \& Fernandez 1996). This last assumption is more strictly
satisfied by the largest objects, which should, through energy equipartition,
evolve to the lowest eccentricity in the swarm 
(Stewart \& Wetherill 1988; Stern \& Colwell 1997b)\\
Although the main motion of KB objects (KBOs) is Keplerian rotation
around the Sun, the motion is perturbed by encounters with other objects
and by collisions.
Encounters influence the structure of the system
in several ways:\\
~~  a) Relaxation; \\
~~  b) Equipartition;\\ 
~~  c) Escape;\\
~~  d) Inelastic encounters.\\
Each of the quoted effects has greater or smaller importance
in a system evolution
according to the system characteristics. In the case of a system like
KB, inelastic encounters have a fundamental role because 
KBOs have on average a much smaller escape speed than the
rms velocity dispersion, $\sigma$.
If $r_{\ast}$ is the radius of a KBO, $v_{\ast}=\sqrt{2 G m/r_{\ast}}$ is the
escape speed from the KBO surface,  $\Theta =\frac{v^2_{\ast}}{4 \sigma^2}$
is the Safronov number, $n$ the number density and $\sigma$ the velocity
dispersion, then the collision time $ t_{\rm coll}$ for a population of
planetesimals with a Gaussian distribution in dispersion
velocity is given by:
\begin{equation}
t_{\rm coll}=\frac{1}{16 \sqrt{\pi} n \sigma r_{\ast}^2 (1+\Theta)}
\label{eq:tcoll}
\end{equation}
(Binney \& Tremaine 1987; Palmer et al. 1993).\\
Within $1 {\rm AU}$ from the Sun, a population of a few hundred km-sized
planetesimals, with several Earth masses in total, would have
$t_{\rm coll} \simeq 10^4 {\rm yr}$, while at distances $> 50 {\rm AU}$
$t_{\rm coll}$ becomes
comparable with the age of the solar system.
Indeed by means of N-body simulations, 
Stern (1995) showed how collisional
evolution plays an important role through KB and that it
has a dominant role at $r>42 {\rm AU}$. The effect of collisions is that of
inducing energy dissipation in the system but at the same time collisions
are important in the growth of QB1 objects, Pluto-scale and larger
objects starting from 1 to 10 km building blocks in a time that in
some plausible circumstances is as little as $\simeq 100  - 200 {\rm Myr}$.\\
Mutual gravitational scattering induces random velocity in two
different ways: one is viscous stirring which converts solar
gravitational energy into random kinetic energy of planetesimals.
Energy is transferred from circular, co-planar orbits with zero
random velocities to eccentric, mutually inclined orbits with
nonzero random velocities. The other is dynamical friction which transfers
random kinetic energy from the larger planetesimals to the smaller
ones (Stewart \& Wetherill 1988; Ida 1990;
Ida \& Makino 1992; Palmer et al. 1993). Unlike viscous stirring,
the exchange of energy does not depend on the differential
rotation of the mean flow for its existence. Dynamical friction
would drive the system to a state of equipartition of kinetic
energy but viscous stirring opposes this tendency.
In the dispersive regime,
the time scales of stirring and dynamical
friction are almost equal to the two-body relaxation time
\begin{equation}
T_{\rm 2B}\simeq \frac{1}{\pi r^2_{\rm G} n \sigma \ln \Lambda} \nonumber \\
\end{equation}
where $r^2_{\rm G}$ is the
gravitational radius, $n$ the number density, 
\begin{equation}
\Lambda=\frac{b_{\rm max} \overline{|v|^2}}{G(m1+m2)}
\nonumber
\end{equation}
$b_{\rm max}$ being the largest impact parameter and $\overline{|v|^2} $ the mean square velocity
of the objects, which for the typical values of the parameters in the KB
is of the order of $ \approx 10^{11}$ and 
$\log \Lambda \approx 25$. \\
In conclusion the
random velocity of the smaller planetesimals is increased by viscous
stirring while the larger planetesimals suffer dynamical friction
due to smaller planetesimals. 

\section{Dynamical friction in KB}

The equation of motion of a KBO can be written as:
\begin{equation}
{\bf \ddot r}= {\bf F}_{\odot}+ {\bf F}_{\rm planets}+{\bf F}_{\rm tide}+
{\bf F}_{\rm GCM}+{\bf F}_{\rm stars}+{\bf R}+{\bf F}_{\rm other}
\end{equation}
(Wiegert \& Tremaine 1997).
The term  ${\bf F}_{\odot}$ represents the force per unit mass from the
Sun,  ${\bf F}_{\rm planets}$ that from planets, $ {\bf F}_{\rm tide}$ that
from the Galactic tide, ${\bf F}_{\rm GCM}$ that from giant molecular clouds,
${\bf F}_{\rm stars}$ that from passing stars, 
${\bf F}_{\rm other}$ that from other sources
(e.g. non-gravitational forces),  while {\bf R} is the dissipative force
(the sum of accretion and dynamical friction terms - 
see Melita \& Woolfson 1996).
If we consider KBOs at heliocentric distances $>70 {\rm AU}$ then
${\bf F}_{\rm planets}$ may be neglected. We also neglect the effects
of non gravitational forces and the perturbations
from Galactic tide, GCMs or stellar perturbations, which are important only for
objects at heliocentric distances $>>100 {\rm AU}$ (Brunini \& Fernandez 1996).
We assume that the planetesimals travel around the Sun in circular
orbits
and we study the orbital evolution of KBOs after
they reach a mass $>10^{22} {\rm g}$.
Moreover we suppose that the role of collisions for our KBOs at distances
$>70 {\rm AU}$ can be neglected. We know that the role of collisions would
be progressively less important with increasing distance from the Sun
because the collision rate, $n v \sigma'$ ($\sigma'$ is the collision cross
section), decreases
due to the decrease in the
local space number density, $n$ of KBOs and the local average crossing
velocity, $v$, of the target body. As stated previously, using
Eq. (\ref{eq:tcoll}) at distances larger than $50 {\rm AU}$ the collision time,
$t_{\rm coll}$, is of the order of the age of the solar system. Besides, 
the energy damping is not dominated by collisional damping but by
dynamical friction damping; also, artificially increasing the collisional
damping the dynamics of the largest bodies is hardly changed
(Kokubo \& Ida 1998). \\
To take account of dynamical friction we need a suitable formula for
a disk-like structure such as KB.
Following Chandrasekhar
\& von Neumann's (1942) method, the frictional force which is experienced by
a body of mass $m_1$, moving through a homogeneous and isotropic
distribution of lighter particles of mass $m_2$, having a
velocity distribution $n(v_2)$ is given by: 
\begin{equation}
{\bf F}=-4\pi m_1 m_2 (m_1+m_2) G^2 \int _0^{v_1} n(v_2)d v_2 
\frac{{\bf v_1}}{v_1^3}\log \Lambda  
\label{eq:cha}
\end{equation}
(Chandrasekhar 1943); 
where $\log \Lambda $ is the Coulomb logarithm, $m_1$ and $m_2$ 
are, respectively, the masses of the test particle and 
that of the field one, and $v_1$ and $v_2$ the respective velocity, 
$n(v_2) d v_2$ is the number of field particles with velocities 
between $v_2$, $v_2+d v_2$. \\ 
If the velocity distribution is Maxwellian Eq. (\ref{eq:cha}) 
becomes:
\begin{eqnarray}
{\bf F} &= &-4\pi n m_1 (m_1+m_2) \rho G^2  
\frac{{\bf v_1}}{v_1^3}  \cdot \nonumber \\
 &   &
\log \Lambda [erf(X)-2 X \exp(-X^2)/\sqrt(\pi)] 
\label{eq:chac}
\end{eqnarray}  
(Chandrasekhar 1943, Binney \& Tremaine 1987),where  
$\rho$ is the density of field particles and $X=v_1/\sqrt{(2 \sigma)}$, 
$\sigma$ being the velocity dispersion. 
Eq. (\ref{eq:chac}) cannot be used for systems not spherically 
symmetric except for the case of objects moving in the equatorial 
plane of an axisymmetric distribution of matter. These objects, in 
fact, have no way of perceiving that the potential in which they move 
is not spherically symmetric. \\
We know that KB is a disc and consequently for objects 
moving away from the disc plane we need a more general formula than
Eq. (\ref{eq:chac}). Moreover 
dynamical friction
in discs differs from that in spherical isotropic three dimensional
systems. First, in a disc close encounters give a contribution to the friction
that is comparable to that of distant encounters (Donner \& Sundelius 1993; 
Palmer et al. 1993).
Collective effects in a disc are much stronger than in a three-dimensional
system.  
The velocity dispersion of particles in a disc
potential is anisotropic. N-body simulations and observations show that
the radial component of the dispersion, $\sigma_{\rm R}$, and the vertical one,
$\sigma_{\rm z}$, are characterized by a ratio $\sigma_{\rm R}/\sigma_{\rm z} \simeq 0.5$
for planetesimals
in a Keplerian disc (Ida et al. 1993). The velocity dispersion evolves through
gravitational scattering between particles. Gravitational
scattering between particles transfers the energy of the systematic rotation
to the random motion (Stewart \& Wetherill 1988).
In other words the velocity distribution of a Keplerian particle disc is
ellipsoidal with ratio 2:1 between the radial and orthogonal ({\it z}) directions
(Stewart \& Wetherill 1988).
According with what previously told, we assume
that the matter-distribution is disc-shaped, having a velocity 
distribution:
\begin{equation}
n({\bf v},{\bf x})=n({\bf x})\left( \frac 1{2\pi }\right)
^{3/2}\exp \left[ -\left( \frac{v_{\parallel }^2}{2\sigma _{\parallel }^2}+%
\frac{v_{\perp }^2}{2\sigma _{\perp }^2}\right) \right] \frac 1{\sigma
_{\parallel }^2\sigma _{\perp }}
\end{equation}
(Hornung \& al. 1985, Stewart \& Wetherill 1988) 
where $ v_{\parallel }$ and $\sigma_{\parallel}$ are the velocity and the velocity dispersion in the 
direction parallel to the plane while  $ v_{\perp }$ and
$\sigma_{\perp}$ are the same in the 
perpendicular direction. We suppose that  $\sigma_{\parallel}$ 
and $\sigma_{\perp}$ are constants and their ratio is simply taken 
to be 2:1.
Then according to Chandrasekhar (1968) and Binney (1977) we may 
write the force components as:
\begin{eqnarray}
F_{\parallel } &= & k_{\parallel}v_{1 \parallel}= B_{\parallel }v_{1\parallel } 
\cdot \nonumber \\
 &   &
\left[ 2\sqrt{2\pi } \overline{n} G^2\lg \Lambda  m_1m_2\left(
m_1+m_2\right) 
\sqrt{1-e^2}\frac 1{\sigma _{\parallel }^2\sigma _{\perp
}}\right]\label{eq:b1}
\end{eqnarray}
\begin{eqnarray}
F_{\perp } &= &k_{\perp}v_{1 \perp}= B_{\perp
}v_{1_{\perp }}  \cdot \nonumber \\
 &   &
 \left[ 2\sqrt{2\pi } \overline{n} G^2\lg \Lambda m_1m_2\left( m_1+m_2\right) 
\sqrt{1-e^2}\frac 1{\sigma _{\parallel }^2\sigma _{\perp }}\right]\label{eq:b2}
\end{eqnarray}
where
\begin{eqnarray}
B_{\parallel } &= &\int_0^\infty dq\exp \left[ -\frac{v_{1\parallel }^2}{%
2\sigma _{\parallel }^2}\frac 1{1+q}-\frac{v_{1\perp }^2}{2\sigma
_{\parallel }^2}\frac 1{1-e^2+q}\right]  \cdot \nonumber\\
 &   &
\frac 1{\left[ \left( 1+q\right)
^2\left( 1-e^2+q\right) ^{1/2}\right] }
\label{eq:b3}
\end{eqnarray}
\begin{eqnarray}
B_{\perp } &= &\int_0^\infty dq\exp \left[ -\frac{v_{1\parallel }^2}{2\sigma
_{\parallel }^2}\frac 1{1+q}-\frac{v_{1\perp }^2}{2\sigma _{\parallel }^2}%
\frac 1{1-e^2+q}\right]  \cdot \nonumber \\
 &   &
\frac 1{\left[ \left( 1+q\right) ^2\left(
1-e^2+q\right) ^{3/2}\right] }
\label{eq:b4}
\end{eqnarray}
and
\begin{equation} 
e=\sqrt{(1-\frac{\sigma_{\perp}^2}{\sigma_{\parallel}^2})}
\end{equation}
while
$\overline{n}$ is the average spatial density.
When $B_{\perp}>B_{\parallel}$ the drag caused by dynamical friction 
will tend to increase the anisotropy of the velocity distribution of 
the test particles. The frictional drag on the test particles may be 
written:
\begin{equation}
{\bf F}=-k_{\parallel}v_{1 \parallel} {\bf e_{\parallel}}-
k_{\perp}v_{1 \perp}{\bf e_{\perp}}
\label{eq:dinn}
\end{equation} 
where ${\bf e_{\parallel}}$ and ${\bf e_{\perp}}$ are two unity vectors 
parallel and perpendicular to the disc plane.\\
This result differs from the classical Chandrasekhar (1943) formula.
Chandrasekhar's result tells that dynamical friction force, ${\bf F}$,
is always
directed as $-{\bf v}$. This means that if a massive body moves,
for example, in a
disc in a plane different from the symmetry plane,
dynamical friction causes it to spiral through
the center of the mass distribution always remaning on its own plane. It
shall reach the disc plane only when it reaches the centre of the
distribution.
Eq. (\ref{eq:dinn}) shows that
a massive object suffers drags 
$-k_{\parallel}v_{1 \parallel}$ and $-k_{\perp}v_{1 \perp}$ 
in the directions
within and perpendicular to the equatorial plane of the distribution.
This means that the object shall find itself confined to the plane of the
disc before it reaches the centre of the distribution (this means
that also inclinations are damped).
In other words the dynamical drag experienced by an object of 
mass $m_1$ moving through a non-spherical distribution of
less massive objects
of mass $m_2$ is not directed in the direction of the
relative movement of the massive particle and the centre of mass of the less
massive objects (as
in the case of spherically symmetric distribution of matter).
As a consequence the already flat distribution of more massive 
objects will be 
further flattened during the evolution of the system (Binney 1977).
The objects lying in the plane, as previously told, have no way 
to perceive that they are moving in a not spherically symmetric 
potential. Hence we expect that the dynamical drag is directed 
in the direction opposite to the motion of the particle:
\begin{equation}
{\bf F} \simeq -k_{\parallel}v_{\parallel} {\bf e_{\parallel}}
\end{equation} 

\section{Parameters used in the simulation}

To calculate the effects of dynamical friction, introduced in the previous
section, on a disc-like structure as KB, we cannot use the classical
Chandrasekhar's (1943) theory but we need equations specified for distributions
like KB. These equations were written in the previous section
(Eqs. \ref{eq:b1} $ -  $ \ref{eq:b4}).
To calculate the effect of dynamical friction on the orbital evolution of
the largest bodies 
we suppose that $\sigma_{\perp}$ and
$\sigma_{\parallel}$ are constants and that
$\sigma_{\parallel}$=$2 \sigma_{\perp}$. We need also the mass density
distribution in the disc. We assume a heliocentric ($R$) distribution in
surface mass density $\Sigma \propto R^{-2}$ and a total primordial mass,
$M=50 M_{\oplus}$ in the region $30  - 50 {\rm AU}$ (Stern \& Colwell 1997b).
To evaluate the dynamical friction force we need the spatial distribution
of the field objects, $\overline{n}$. To reproduce the quoted
surface density we need a number density decreasing as a function of distance
from the Sun, $R$, to the third power (Levison \& Duncan 1990):
\begin{equation}
n=n_o R^{-3}
\label{eq:lev}
\end{equation}
We remember also that the KB is a disc and then we use the mass
distribution given by a Myiamoto-Nagai disc (Binney \& Tremaine 1987; 
Wiegert \& Tremaine 1997):
\begin{eqnarray}
n(R,z) &= & \left( \frac{b^2M}{4\pi }\right)  \cdot \nonumber \\
 &   &
\frac{aR^2+\left( a+3\sqrt{%
z^2+b^2}\right) \left( a+\sqrt{z^2+b^2}\right) ^2}{\left[ R^2+\left( a+\sqrt{%
z^2+b^2}\right) ^2\right] ^{5/2}\left( z^2+b^2\right) ^{3/2}}
\end{eqnarray}
which in the disc plane reproduces the quoted surface density and
Eq. (\ref{eq:lev}) $[n(R,0) \propto R^{-3}]$.
Here $M$ is the disc mass, $a$ and $b$ are parameters describing the
disc characteristic radius and thickness.
Because there is presently
no information on the way in which ensemble-averaged inclinations ($\langle i
\rangle $)
and eccentricities ($\langle e
\rangle$) vary in the KB, we adopt a disc with a width 
$\langle i
\rangle = \frac{1}{2}\langle e
\rangle$ (Stern 1996b).
The equations of motion were integrated in heliocentric coordinates
using the Bulirsch-Stoer method.

\section{Results}

The model described was integrated for several values of masses, starting from
$m=10^{22} {\rm g}$, supposing the KBOs move on circular orbits. We studied
the motion of KBOs both on inclined orbits, in order to study the evolution
of the inclination with time, and on orbits on the plane of the disc, in order
to study the drift of the KBOs from their initial position. The masses
of the KBOs, $M$, were considered constant during the whole integration in order
to reduce the number of differential equations to solve. Moreover we use 
$M<< N m$ and $m<< M$, $N$ and $m$ being the total number and the
mass of the swarm of field particles in which the KBO moves. The
assumption that field particles have all equal masses, $m$, does not affect
the results, since dynamical friction does not depend on the individual
masses of these particles but on their overall density. 
The results of our calculations are shown in Figs. 1 $ -  $ 4.\\
In Fig. 1 we plot the values of $\langle z
\rangle/z_o$ versus time for planetesimals
having masses $10^{22} {\rm g}$ and $10^{23} {\rm g}$. The {\it z} coordinate is orthogonal to
the plane of the disc and $z_o=7 {\rm AU}$ while $t_{\rm o}=10^8 {\rm yr}$.   
The brackets $\langle 
\rangle$ are mean values obtained averaging over a suitable number
of orbital oscillations. As shown in the figure the decay of the inclination
of the planetesimal of $10^{22} {\rm g}$ (dotted line) has a timescale larger than the
age of the solar system and on the order of $3 \times 10^{10} {\rm yr}$. The
inclination of the more massive planetesimal (full line) decays in a time
$\simeq 3 \times 10^9 {\rm yr}$. This is due to the fact that dynamical friction
effects increase with the mass $M$ of the KBO. In fact (see Fig. 2)
when the mass of the
KBO is $10^{25} {\rm g}$ the decay time reduces to $3 \times 10^8 {\rm yr}$. 
Dynamical friction makes the orbits of KBOs undergo some collimation along the {\it z} direction, 
characterized by a low value of the dispersion velocity. This was expected
because 
it is known that the larger the velocity dispersion along the direction
of motion, the lesser the effect of dynamical friction (Pesce et al. 1992).
\begin{figure}
\psfig{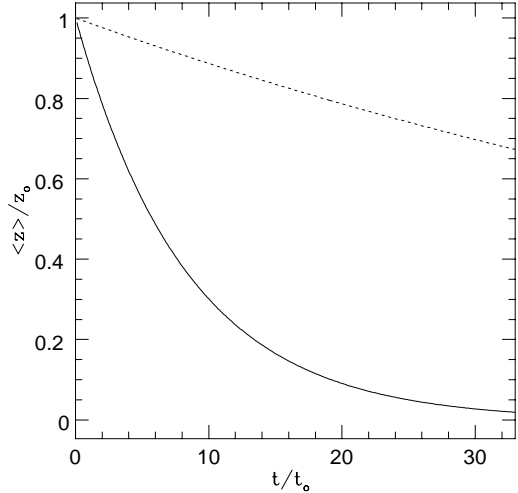}
\caption[]{Height from the plane versus time for a planetesimal of
$10^{22} {\rm g}$, dotted line,
and one of $10^{23} {\rm g}$, full line.
Time is measured in units of $t_{\rm o}=10^8 {\rm yr}$  while height
is measured in $z_o=7{\rm AU}$. }
\label{Fig. 1}
\end{figure}
Binney (1977) has found an efficient collimation of orbits along the main axis 
of the velocity dispersion tensor in the case of an anisotropic 
axisymmetric system, in which the principal velocity dispersions 
have constant values. \\
The result obtained is in agreement with previous studies
of the damping of the inclinations of very massive
objects by Ida (1990), Ida \& Makino (1992) and Melita \& Woolfson (1996).
In fact in the semi-analytical theory by Ida (1990) the timescale for
inclinations damping due to dynamical friction is almost equal to the
two-body relaxation time
\begin{equation}
T_{\rm damp} \simeq T_{\rm 2B} \simeq \frac{1}{n G^2 M^2\ ln\Lambda/\sigma^3} \simeq
\frac{1}{\frac{\Sigma}{m} \frac{G^2 M^2}{\sigma^4} \Omega \ln \Lambda}
\end{equation}
where $\Sigma$ is the surface density of small objects and $\Omega$ is the
Keplerian frequency. The timescales given by Ida's theory
for our planetesimals
of $10^{22} {\rm g}$,  $10^{23} {\rm g}$,  $10^{25} {\rm g}$
are respectively $4 \times 10^{10} {\rm yr}$,
$4 \times 10^{9} {\rm yr}$, $4 \times 10^{8} {\rm yr}$, 
in agreement with our results. \\
\begin{figure}
\psfig{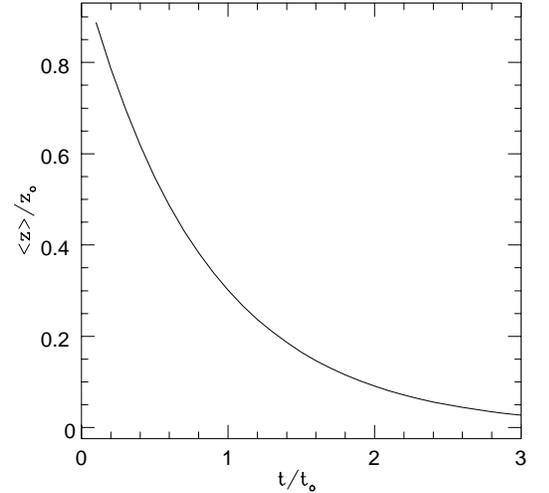}
\caption[]{Height from the plane versus time for a planetesimal of $10^{25} {\rm g}$.  
Time is measured in units of $t_{\rm o}=10^8 {\rm yr}$  while height 
is measured in $z_o=7{\rm AU}$. }
\label{Fig. 2}
\end{figure}
\begin{figure}
\psfig{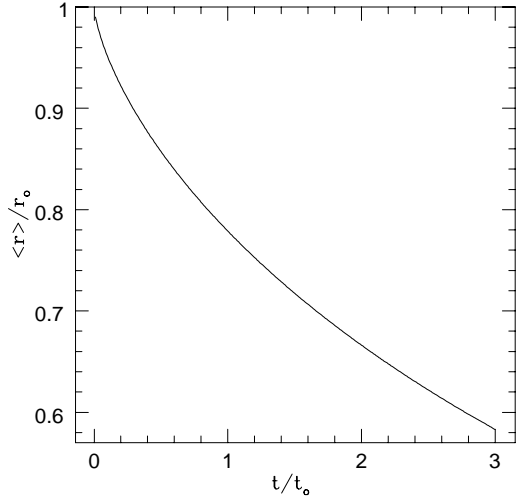}
\caption[]{Heliocentric distances in the disk plane for a 
planetesimal of $10^{23}g$. 
Time is measured in units of $t_{\rm o}=10^{10} {\rm yr}$  while distances  
are measured in $r_{\rm o}=70{\rm AU}$. }
\label{Fig. 3}
\end{figure}
The effect of dynamical friction on the
semi-major axis is plotted in Fig. 3 and Fig. 4. In Fig. 3
we plot $\langle r
\rangle/r_{\rm o}$ versus time for a planetesimal of mass $10^{23} {\rm g}$.
Here $r$ is the in-plane radial heliocentric
distance of the planetesimal while $r_{\rm o}=70 {\rm AU}$ and
$t_{\rm o}=10^{10} {\rm yr}$. As shown, the time required
to a planetesimal of the quoted mass to reach $40 {\rm AU}$ is
$\simeq 3 \times 10^{10} {\rm yr}$, which is an order of magnitude larger than
the damping timescale. This is due, in agreement with what previously told,
to the fact that in the plane the dispersion velocity is larger than that
in the $z$ direction. Increasing the mass to
$10^{25} {\rm g}$ the time needed
for a planetesimal to reach $40 {\rm AU}$ decreases to $4 \times 10^8 {\rm yr}$
(here $t_{\rm o}=10^6 {\rm yr}$)
(see Fig. 4). The threshold planetesimal mass, $M_{\rm threshold}$,
that starts orbital migration is
$\sim 10^{24}$ g. This mass scales with the disk density as:
\begin{equation}
M_{\rm threshold} \sim \frac{10^{24}}{(\rho/{\rm 10^{-16} g/cm^3})} g
\end{equation}
We recall that we do not take into account the effects
of the planets because we considered planetesimals initially at distances
$>70 {\rm AU}$, but when the planetesimal moves towards the region of influence
of planets the role of these must be taken into account. \\
We have some difficulty to compare this result with previous studies
because the problem of the decay of the semi-major axis has
not been particularly studied.
So far, many people have assumed a priori that radial migration
due to dynamical friction is much slower than damping of velocity dispersion 
due to dynamical friction. Therefore most of studies of
dynamical friction were concerned only
with damping of velocity dispersion (damping of the eccentricity, $e$,
and inclination, $i$), adopting local coordinates.
Analytical work by Stewart \& Wetherill (1988) and by Ida (1990), 
adopted local coordinates. N-body simulation
by Ida \& Makino (1992) adopted non-local coordinates, but did not
investigate radial migration. Only the density wave approach
by Goldreich \& Tremaine (1979, 1980) and Ward (1986) considered
radial migration. However, the relation of this approach to
particle orbit approach is not clear. Furthermore, a few numerical simulation
has been devoted to investigate radial migration. \\
In any case we shall compare our result with that by
Ward (1986) supposing that this density approach
describes correctly the radial migration. Following
Ward (1986) the characteristic orbital decay time 
of a disc perturber
is given by:
\begin{equation}
T_{\rm dec} = \frac{1}{\Omega |C|\mu} \frac{M_{\odot}}{\Sigma a^2}
\left(\frac{c}{a \Omega}\right)^2
\label{eq:ward}
\end{equation}
where $\mu=M/M_{\odot}$, and the nondimensional factor $C$, depending on the
disc's surface density gradient, $k$, and the adiabatic index, $s$, for
a disc with $Q=\infty$ is $\simeq -18$, $c$ 
is the gas sound speed (in a planetesimals system this must be replaced
by the velocity dispersion). The timescale for the perturber
to drift out of a region of radius $r$ is given by (Ward 1986):
\begin{equation}
T_{drift} \simeq \frac{c}{r \Omega} T_{dec} 
\end{equation}
Using $\Omega=\sqrt{G M_{\odot}/r^3}$, calculating $\Sigma$ supposing that 
the disc mass is uniformly spread in the region $40  -  70 {\rm AU}$ and
$ c = 20000 \, {\rm cm/s}$ we find  $T_{\rm drift} \simeq 7 \times 10^8 {\rm yr} $
for a
planetesimal having $M=10^{25} {\rm g}$, a little larger than the value
previously found. We have to remember that Ward's (1986) model
supposes that the solar nebula is two-dimensional and that in a finite thickness
disc other damping mechanisms may come into play perhaps invalidating
Ward's result.  \\
\begin{figure}
\psfig{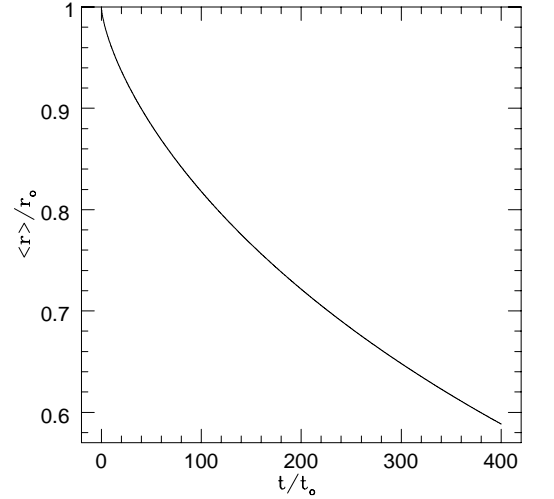}
\caption[]{Heliocentric distances in the disk plane for a 
planetesimal of $10^{25} {\rm g}$. 
Time is measured in units of $t_{\rm o}=10^6 {\rm yr}$  while distances  
are measured in $r_{\rm o}=70{\rm AU}$. }
\label{Fig. 4}
\end{figure}
This last calculation shows that
in a timescale less than the age of
the solar system,
objects of the mass of Pluto
$\simeq 0.002 M_{\oplus}$ may move
from the region $>50 {\rm AU}$ towards the position actually
occupied by the planet. This opens a third possibility to the standard
scenarios for Pluto formation (in situ formation,
formation at $20 - 30  {\rm AU}$
and transport outwards) namely that Pluto was created beyond $~50 {\rm AU}$,
and then transported inwards.
One of the problems of Pluto formation
in the first two scenarios is that the growth of Neptune caused accretion
to be inhibited and then it is necessary that the time for 
accreting an object of Pluto mass was shorter than the timescale of Neptune
formation. This problem is not present in the third scenario
because at heliocentric distances larger than
$50 {\rm AU}$ Neptune never induced significant eccentricities on most orbits in the
region. Hence the dynamical conditions necessary for growth may have persisted
for the whole age of the solar system
and consequently Pluto could have formed later than
in the other two quoted scenarios.
The mechanism responsible for the transport of Pluto from the quoted region
to that nowadays occupied might have been dynamical friction. \\
Three possible objections to this last model are:\\
1) Moving to greater and greater distances both the disc density and the
velocity dispersions decrease and consequently the accretion
times increase. Can a
Pluto-scale body form at distances of 70 AU?\\
2) A possible explanation for Pluto's orbital parameters is connected to
outwards migration of Neptune (Malhotra 1993).
If this works for Neptune, could it also work for Pluto-like
objects in the KB? How much might Pluto have moved out? Could that
compensate for the effect of the frictional motion? \\
3) How can one explain the odd orbital parameters of Pluto's orbit
($e=0.25$, $i=17$ degrees to the ecliptic)?\\
Surely, as made clear by the first objection, formation of large bodies
is more and more difficult moving away from the inner parts of Solar system.
Although growth times at 70 ~AU are about 4-5 times longer than at 40 AU,
Pluto-mass bodies can indeed be grown at this distance, from 1 to 10 km
building blocks, in $\sim 1 {\rm Gyr}$, if the mean disc eccentricity,
$<e> \simeq 0.001$, and if the KB mass interior to 50 AU was, as previously
stated in Sect. ~4, 30-50 $M_{\oplus}$ and continued outwards with
$\Sigma \propto R^{-2}$ (see Stern \& Colwell 1997b).\\ 
The answer to the second objection is the following. \\
Orbital migration of a planet can be accomplished through two mechanisms.
In the first mechanism a planet and the circumstellar disc interact tidally
which results in angular momentum transfer between the disc and the planet
(e.g. Goldreich \& Tremaine 1980; Ward 1997). The planet's motion in the
disc excites density waves both interior and exterior to the planet.
If the planet is large enough (at least several Earth masses), it is
able to open and sustain a gap. It establishes a barrier to any radial
disc flow due to viscous diffusion and it becomes locked to the disc and
must ultimately share its fate (this is known as $type$ II drift).
In this case both inwards and outwards planet migration are allowed.
In fact in a viscous disc, gas inside the radius of maximum viscous stress,
$r_{\rm mvs}$, drifts inwards as it loses angular momentum while gas outside
$r_{\rm mvs}$ expands outwards as it receives angular momentum
(Lynden-Bell \& Pringle 1974). Neptune's outwards migration is due to
the fact that the gas in the Neptune forming region has a tendency to
migrate outwards (Ruden \& Lin 1986).\\
If the planet is not able to sustain a gap, the net torque from the disc
is still not zero and it migrates inwards in a shorter timescale
($type$ I drift).\\
Pluto being a low mass planet it can migrate only by means of $type$ I
drift, this means that it can only migrate inwards.\\
In the second mechanism a planet can undergo orbital migration as a
consequence of gravitational scattering between itself and residuals
planetesimals. If a planetesimal in a near-circular orbit similar to that of
the planet is ejected into a Solar system escape orbit, the planet suffers
a loss of orbital angular momentum and a corresponding change of
orbital radius. Conversely, planetesimals scattered inwards would cause
an increase of orbital radius and angular momentum of the planet.
A single massive planet scattering a population of planetesimals
in near-circular orbits in the vicinity of its own orbit would suffer
no net change of orbital radius as it scatters approximately
equal numbers of planetesimals inwards and outwards. However in
some peculiar situations, such as that encountered in the region
of Jovian planets,
things go differently from this picture
(Fernandez \& Ip 1984). In particular, as Jupiter preferentially
removes the inwards scattered Neptune planetesimals, the planetesimal
population encountering Neptune at later times is increasingly
biased towards objects with specific angular momentum larger than Neptune's.
Encounters with this planetesimal population produce a net gain of angular
momentum, hence an increase in its orbital radius. Evidently this situation
is not the one present in the outer Solar system region, occupied by Pluto
in our model. In other words, there is no reason to suppose that Pluto
moved outwards like Neptune. \\
For what concerns the third objection, a possible answer is that Pluto
gained high eccentricity and inclination in a similar way
to that described by Malhotra (1993, 1995b). There has, of course, been much
speculation as to the
origin of the extraordinary orbit of Pluto (Lyttleton 1936;
Farinella et al. 1979; Olsson-Steel 1988;
Malhotra 1993). All but one (Malhotra (1993, 1995b))
of these speculations require one or more low-probability
"catastrophic"  events. In Malhotra's (1993, 1995b) model, Neptune's orbit
may have expanded considerably, and its exterior orbital resonances
would have swept through a large region of trans-Neptunian space. During
these resonances sweeping, Pluto could have been captured into the 3:2
orbital period resonance with Neptune and its eccentricity and inclination
would have been pumped up during the subsequent evolution. \\
The phenomenon
of capture into resonance as result of some dissipative forces is common
in nature. Weidenschilling \& Davies (1985) studied resonance trapping of
planetesimals by a protoplanet in association with gas drag. Many of the
characteristics of this effect have been studied. Patterson (1987) and
Beauge et al. (1994) have investigated its cosmogonic implications. The
stability of the orbits and capture probabilities have been studied by
Beauge \& Ferraz-Mello (1993), Gomes (1995). Melita \& Woolfson (1996)
showed that a three-body system (Sun and two planets) under the influence
of both accretion and dynamical friction forces, evolve into planetary
resonance when the inner body is more massive. In general capture into
a stable orbit-orbit resonance is possible when the orbits of two bodies
approach each other as a result of the action of some dissipative
process. The transition from a non-resonant to a resonant orbit
depends sensitively upon initial conditions and the rate of orbital evolution
due to the dissipative effects. Borderies \& Goldreich (1984) showed that
for single resonance and in the limit of slow "adiabatic" orbit evolution,
the probability of capture for the 3:2 Neptune resonance is 100\% for initial
eccentricity less than $\sim 0.03$ and reduces to less than 10\% for
initial eccentricities exceeding 0.15 (see Malhotra 1995b).\\
In our model Pluto reaches its actual position in a time ($>10^8$ yr)
larger than Neptune's orbital migration timescale ($10^6-10^7$ yr;
see Ida et al. 1999). Then Neptune was in its actual position when Pluto
reached its own. Pluto, before the resonance encounter, Pluto has an initial low
eccentricity $\sim 0.03$, because of eccentricity and inclination
damping due to dynamical friction, and its migration is very slow
($> 10^8$ yr).
The capture probability is then very high. The increase in eccentricity and
inclination is naturally explained by Malhotra's theory.

\section{Conclusions}

In this paper we studied how dynamical friction due to small
planetesimals influence the
evolution of KBOs having masses larger than
$10^{22} {\rm g}$. We find that mean eccentricity of large mass particles
is reduced by dynamical friction by small mass particles in timescales
shorter than the age of the solar system for objects of mass
equal or larger than $10^{23} {\rm g}$. We also studied the effect of dynamical
friction on the evolution of the semi-major axis of the largest
planetesimals. We find that even if dynamical friction is less effective
in transferring planetesimals towards the inner part of the solar system,
with respect to the damping of inclinations that it is able to produce,
the timescale for radial migration is shorter than the age of the solar
system for large enough masses ($ \geq 10^{24} {\rm g}$).\\
Finally, our calculation show the dynamical friction may be the mechanism 
responsible for the transport of objects like Pluto from regions with $ r > 50 {\rm AU}$ towards the position nowdays occupied and this opens a third possibility 
for Pluto formation that eliminates the problem of the Neptune formation.

\begin{acknowledgements}
We would like to thank the referee  C. Dominik for helpful comments
which led to improve our paper. Besides we are grateful to S. Ida 
for helpful and stimulating discussions during the period in which this 
work was performed; to A. Stern for interesting informations and to 
E. Recami for some useful comments.
\end{acknowledgements}


\begin{thebibliography}{}
\bibitem{} Aumann H.H., Gillette F.C., Beichman C.A., et al., 1984, ApJ 278, L23
\bibitem{} Beauge C., Ferraz-Mello S., 1993, Icarus 103, 301
\bibitem{} Beauge C., Aarseth S.J., Ferraz-Mello S., 1994, MNRAS 270, 21
\bibitem{} Beckwith G., Sargent A.I., 1993, Protostars and Planets III, Levy  E.H., Lunine J.I., (eds.) The University of Arizona Press. Tucson, p. 521
\bibitem{} Brown M.E., Kulkarni S.R., Ligget T.J, 1997, ApJ 490, L119
\bibitem{} Brunini A., Fernandez J.A., 1996, A\&A 308, 988
\bibitem{} Binney J., 1977, MNRAS 181, 735
\bibitem{} Binney J., Tremaine S., 1987, Galactic Dynamics, in Princeton Series in Astrophysics, Princeton University Press.
\bibitem{} Borderies N., Goldreich P., 1984, Cel. Mech. 32, 127
\bibitem{cha2} Chandrasekhar S., von Neumann J., 1942, ApJ 95, 489
\bibitem{cha1} Chandrasekhar S., 1943, ApJ 97, 255
\bibitem{} Chandrasekhar S., 1968, Ellipsoidal figures of equilibrium, Yale University Press.
\bibitem{} Cochran A.L., Levison H.F., Stern S.A., Duncan M.J., 1995, AJ 455, 342
\bibitem{} Cochran A.L., Levison H.F, Tamblyn P., Stern S.A., Duncan M.J., 1998, ApJ 503, L89 (see also SISSA preprint astro-ph/9806210)
\bibitem{} Donner K.J., Sundelius B., 1993, MNRAS 265, 88
\bibitem{} Duncan M.J., Quinn T., Tremaine S., 1988, ApJ 328, L69
\bibitem{} Duncan M.J., Levion H.F., Budd S.M., 1995, AJ 110, 373
\bibitem{} Edgeworth K.E., 1949, MNRAS 109, 600
\bibitem{} Farinella P.A., Milani A., Nobili M., Valsecchi G.B., 1979, The Moon and the Planets 20, 415-421 
\bibitem{} Fernandez J.A., 1980, MNRAS 192, 481
\bibitem{} Fernandez J.A., Ip W.H., 1984, Icarus 58, 109
\bibitem{} Goldreich P., Tremaine S., 1979, ApJ 233, 857
\bibitem{} Goldreich P., Tremaine S., 1980, ApJ 241, 425
\bibitem{} Gomes R.S., 1995, Icarus 115, 47
\bibitem{} Hills J.G., 1981, AJ 86, 1730
\bibitem{} Holman M., Wisdom J., 1993, AJ 105, 1987
\bibitem{} Horedt G.P., 1985, Icarus 64, 448
\bibitem{} Hornung P., Pellat R., Borge P., 1985, Icarus 64, 295
\bibitem{} Ida S., 1990, Icarus 88, 129
\bibitem{} Ida S., Makino J., 1992, Icarus 98, 28
\bibitem{} Ida S., Kokubo E., Makino J., 1993, MNRAS 263, 875
\bibitem{} Ida S., Bryden G., Lin D.N.C., Tanaka H., 1999, submitted to ApJ, (preprint available from http://www.geo.titech.ac.jp/nakazawalab/ida/DG.html)
\bibitem{} Jewitt D., Luu J.X., 1993, Nat. 362, 730
\bibitem{} Jewitt D., Luu J.X., 1995, AJ 109, 1867
\bibitem{} Kokubo E., Ida S., 1998, Icarus 131, 171
\bibitem{} Kuiper G. P., 1951, in Astrophysics: A Topical Symposium, edited by J. A. Hynek (McGraw Hill, New York), pp. 357-424
\bibitem{} Levison H.F., 1996, in ASP Conf. Proc. 107, Completing the Inventory of the Solar System, ed. T.W. Rettig \& J.M. Hahn (San Francisco: ASP), p. 173
\bibitem{} Levison H.F., Duncan M.J., 1990, AJ 100, 1669
\bibitem{} Levison H.F., Duncan M.J., 1993, ApJ 406, L35
\bibitem{} Linden-Bell D., Pringle J.E., 1974, MNRAS 168, 603
\bibitem{} Luu, J., Marsden B.G., Jewitt D., et al., 1997, Nat. 387, 573
\bibitem{} Lyttleton R.A., 1936, MNRAS 97, 108
\bibitem{} Malhotra R., 1993, Nat. 365, 819
\bibitem{} Malhotra R., 1995a, AJ 110, 420
\bibitem{} Malhotra R., 1995b, SISSA preprint astro-ph/9504036
\bibitem{} Melita M.D., Woolfson M.M., 1996, MNRAS 280, 854
\bibitem{} Olsson-Steel D.I, 1988, A\&A 195, 327
\bibitem{} Oort J.H., 1950, Bull. Astr. Inst. Netherlands, 11, 91
\bibitem{} Palmer P., Lin D.N.C., Aarseth S.J., 1993, ApJ 403, 336 
\bibitem{} Patteron C.W., 1987, Icarus 70, 319
\bibitem{} Pesce E., Capuzzo-Dolcetta R., Vietri M., 1992, MNRAS 254, 466
\bibitem{} Quinn T.R., Tremaine S., Duncan M.J., 1990, ApJ 355, 667
\bibitem{} Ruden S.P, Lin D.N.C., 1986, ApJ 308, 883
\bibitem{} Smith B.A., Terrile R.J., 1984, Sci, 226, 1421
\bibitem{} Stern S.A., 1995, AJ 110, 856
\bibitem{} Stern S.A., 1996a, A\&A 310, 999
\bibitem{} Stern S.A., 1996b, AJ 112, 1203
\bibitem{} Stern S.A., Colwell J.E., 1997a, ApJ 490, 879
\bibitem{} Stern S.A., Colwell J.E., 1997b, AJ 114, 841
\bibitem{} Stewart G.R., Kaula W.M., 1980, Icarus 44, 154
\bibitem{} Stewart G.R., Wetherill G.W., 1988, Icarus 74, 542
\bibitem{} Tremaine S., 1990, In Baryonic Dark Matter, Lynden-Bell D., Gilmore G., (eds), Kluwer Dordrecht, p. 37
\bibitem{} Ward W.R., 1986, Icarus 67, 164
\bibitem{} Ward W.R., 1997, ApJ 482, L211
\bibitem{} Weidenschilling S.J., Davies D.R., 1985, Icarus 62, 16
\bibitem{} Weidenschilling S.J., Spaute D., Davis D.R., Marzari F., Ohtsuky K.,  1997, Icarus 128, 429
\bibitem{} Weissmann P.R., 1995, ARA\&A 33, 327
\bibitem{} Weissmann P.R., Levison H.F., 1997, in Pluto and Charon, Stern S.A., Tholen D.J, (eds.), (University of Arizona Press, Tucson) (in press), 559
\bibitem{} Wiegert P., Tremaine S., 1999, Icarus 137, 84
\bibitem{} Yamamoto T., Mizutani H., Kadota A., 1994, PASJ, L5
\end{thebibliography}
\end{document}